\documentclass[11pt]{article}
\usepackage{showkeys}
\begin{document}
%%%%%%%%%%%%%%%%%%%%
\title{{\bf Lie algebraic noncommuting structures from
reparametrisation symmetry }}
%%%%%%%%%%%%%%%%%%%%
\author{
{\bf Sunandan Gangopadhyay $^{}$\thanks{sunandan@bose.res.in}}\\
S.N.Bose National Centre for Basic Sciences,\\JD Block, Sector III, Salt Lake, Kolkata-700098, India\\[0.3cm]
}
\date{}

\maketitle

\begin{abstract}
\noindent We extend our earlier work of revealing both space-space
and space-time noncommuting structures 
in various models in particle mechanics exhibiting
reparametrisation symmetry. We show explicitly (in contrast
to the earlier results in our paper \cite{sg})
that for some special
choices of the reparametrisation parameter $\epsilon$, one can
obtain space-space noncommuting structures which are Lie-algebraic
in form even in the case of the relativistic free particle.
The connection of these structures with the existing models
in the literature is also briefly discussed.
Further, there exists some values of $\epsilon$ for which the 
noncommutativity in the space-space sector can be made to vanish.
As a matter of internal consistency of our approach, we also study
the angular momentum algebra in details.
\\[0.3cm]
\\[0.3cm]
\end{abstract}

\section{Introduction}
The interest in noncommutative theories has been motivated by the results
in string theory. The early results in this subject \cite{witt}, have been
followed by a vast number of papers dealing with the problem of formulating
a noncommutative (NC) quantum mechanics \cite{muthu}, \cite{hari}, \cite{bc}
and field theory. In this context, it has been observed that
an important role is played by redefinitions or change of variables
providing a map between the commutative and noncommutative structures
\cite{nair}, \cite{stern}, \cite{rabin}, \cite{romero}.
\noindent In a recent paper \cite{sg}, in contrast to the earlier
approaches, we have shown that noncommuting
structures can be obtained for models in particle mechanics with 
reparametrisation symmetry. In general, the associative algebraic structure 
$\mathcal{A}_{x}$ which defines a noncommutative space can be defined
in terms of a set of generators $x^{i}$ and relations 
$\mathcal{R}$.  Some important explicit cases are of the form of a canonical
structure \cite{madore}\footnote{In all these cases the index
$i$ representing the spatial coordinates takes values from
$1$ to $d$.} 
\begin{eqnarray}
\{x^{i}, x^{j}\}=\theta^{ij};\quad
\theta^{ij}\in\mathcal{C},
\label{1}
\end{eqnarray}
a Lie algebraic structure
\begin{eqnarray}
\{x^{i}, x^{j}\}={C^{ij}}_{k}x^{k};\quad
{C^{ij}}_{k}\in\mathcal{C},
\label{2}
\end{eqnarray}
and a quantum space structure \cite{wess}
\begin{eqnarray}
x^{i}x^{j}=q^{-1}{R^{ij}}_{kl}x^{k}x^{l};\quad
{R^{ij}}_{kl}\in\mathcal{C}.
\label{3}
\end{eqnarray}
\noindent Gauge theories have been formulated on each of these 
NC spaces based on the notion of covariant coordinates
and Seiberg-Witten (SW) maps
have been established in all cases in \cite{madore}. Hence,
a thorough understanding
of the above types of NC spaces and their emergence 
is important in its own right, if not essential.
Generation of NC phase-space with Lie algebraic forms
of noncommutativity have also appeared in
\cite{amelino}, \cite{lukierski}, \cite{kowalski}, \cite{subir}, \cite{woro}.
%%%%%%%%%%%%%%%%%%%%%%%%%%%%%%%%%%%%%%%%%%%%%%%%%%%%%%%%

\noindent In \cite{sg}, we exploited the reparametrisation symmetry
of the problem to find a nonstandard gauge in which the space-time and 
space-space coordinates become noncommuting. 
There we presented a definite method of finding
this gauge and also showed that the
change of variables relating the nonstandard and standard gauges is
a gauge transformation.
The structures obtained
were Lie-algebraic in the case of a nonrelativistic free particle,
but not so in 
its relativistic counterpart. In this paper, 
we make use of the change of variables derived in our earlier paper
\cite{sg} to show explicitly that for some special choice of the
reparametrisation parameter, one can obtain noncommuting
space-space structures falling in
the Lie-algebraic category even in the relativistic case.
We emphasize that these Lie-algebraic structures
may be useful in giving explicit forms of the
star products and SW maps (discussed in \cite{madore})
by reading off the structure constants of the algebra.

\noindent Moreover, there exists solutions of $\epsilon$ for which
the noncommutativity between spatial coordinates vanish, but the space-time
algebra still remains noncommutative.
%However, the converse is not true.
%There exists no solutions of $\epsilon$ which keeps the space-space
%algebra noncommuting while making the space-time algebra  vanish.
%This indicates that a noncommuting space-space algebra always
%induces a noncommuting space-time algebra but the converse may not true.
However, we do not get solutions for which the space-time algebra vanishes
while the space-space algebra remains noncommuting.
Finally, in order to show the internal consistency of our analysis, we
study the angular momentum algebra in details.

%%%%%%%%%%%%%%%%%%%%%%%%%%%%%%%%%%%%%%%%%%%%%%%%%
\section{Lie algebraic noncommuting structures : relativistic free particle}
In this section, we first give a brief review of reparametrisation
symmetry exhibited by the free relativistic particle \cite{sg}.
The standard reparametrisation invariant action of a
relativistic free particle which propagates in $d + 1$-dimensional ``target
 spacetime" reads\footnote{$x^{\mu}$ are the space-time coordinates,
${\mu} = 0, 1, ...d$, the dot here denotes
differentiation with respect to the evolution parameter $\tau$,
 and the Minkowski metric is $\eta = diag(-1, 1, ..., 1)$. } 
\begin{equation}
S_{0} = -m\int d\tau \sqrt{-\dot{x}^{2}}.
\label{4}
\end{equation}
The canonically conjugate momenta to $x^{\mu}$ are given by
\begin{equation}
p_{\mu} = \frac{m\dot{x}_{\mu}}{\sqrt{-\dot{x}^{2}}}.
\label{5}
\end{equation}
These are subject to the Einstein constraint
\begin{equation}
\phi_{1} = p^{2} + m^{2} \approx 0
\label{6}
\end{equation}
and satisfy the standard PB relations
\begin{equation}
\{x^{\mu}, p_{\nu}\} = \delta^{\mu}_{\nu};  \qquad \{x^{\mu}, x^{\nu}\}
 = \{p^{\mu}, p^{\nu}\} = 0.
\label{7}
\end{equation}
Now using the reparametrisation symmetry of the problem 
(under which the action (\ref{4}) is invariant)
and the fact that $x^{\mu}(\tau)$ transforms as a  scalar
under world-line reparametrisation\footnote{The generator
of the above reparametrisation invariance is proportional to
the Einstein constraint $\phi_{1}$. This
has been discussed in \cite{sg} and we shall not elaborate
on this aspect here.}
 
\begin{eqnarray}
\tau \rightarrow \tau^{'} &=& \tau^{'}(\tau)\nonumber\\
x^{\mu}(\tau)\rightarrow x^{'\mu}(\tau^{'}) &=& x^{\mu}(\tau)
\label{8}
\end{eqnarray}
leads to the following infinitesimal transformation of the space-time
coordinate
\begin{equation}
\delta x^{\mu}(\tau) = x^{'\mu}(\tau) - x^{\mu}(\tau)
= \epsilon \frac{dx^\mu}{d\tau}. 
\label{9}
\end{equation}
The simplest gauge condition to get rid of the gauge freedom generated
by $\phi_{1}$ (\ref{6}), is obtained by 
identifying the time coordinate $x^{0}$ with the evolution parameter $\tau$, 
\begin{equation}
\phi_{2} = x^{0} - \tau \approx 0
\label{9.1}.
\end{equation}
The constraints (\ref{6}, \ref{9.1}) form a second class set with
\begin{equation}
\{\phi_{a}, \phi_{b}\} = 2p_{0}\epsilon_{ab};\quad (a, b=1,2).
\label{9.2}
\end{equation}
The resulting non-vanishing Dirac brackets (DB) are\footnote{The
Dirac brackets are defined as 
$\{A, B\}_{DB} = \{A, B\} - \{A, \phi_{a}\}(\phi^{-1})_{ab}\{\phi_{b}, B\}$, 
where $A$, $B$ are any pair of phase-space variables \cite{dirac}.}:
\begin{eqnarray}
\{x^{i}, p_{0}\}_{DB} = \frac{p^{i}}{p_{0}} \qquad\{x^{i}, p_{j}\}_{DB} = {\delta^{i}}_{j}
\label{9.3}
\end{eqnarray}
which imposes the constraints $\phi_{1}$ and $\phi_{2}$ strongly.

\noindent Using (\ref{9}), the transformations that relates
the primed coordinates in terms
of the unprimed coordinates can be written down in terms
of phase-space variables as:
\begin{eqnarray}
x^{'0} = x^{0} + \epsilon \qquad;
\qquad x^{'i} &=& x^{i} - \epsilon\frac{p^{i}}{p_{0}} 
\label{10}
\end{eqnarray}
where we have used the relation $\frac{dx^{i}}{d\tau}=-\frac{p^{i}}{p_{0}}$.
With the above change of variables (derived from reparametrisation
symmetry) at our disposal, enables us to choose some special
values of the reparametrisation parameter $\epsilon$ which 
leads to noncommuting structures falling in the Lie- algebraic
category (\ref{2}) as we shall see subsequently.

\noindent Setting\footnote{The parameter $\epsilon$ 
of the reparametrisation transformation in (\ref{10a}) is not a Lorentz
scalar (or four-vector). However, this problem is not serious
and we shall present a detailed discussion of this issue later
in the paper.} 
\begin{eqnarray}
\epsilon=-\theta^{0k}p_{k}\frac{p_{0}}{m}
\label{10a}
\end{eqnarray}
and using (\ref{9.3}) and (\ref{10}), we obtain the following 
algebra between the primed coordinates\footnote{Note that $p_{\mu}$'(s) are
gauge invariant objects as $\{p_{\mu}, \phi\}=0$, hence 
$p^{\prime}_{\mu}=p_{\mu}$.}:
\begin{eqnarray}
\{x^{'i}, x^{'j}\}_{DB} = \frac{1}{m}\left(\theta^{0i}p^{j}
 - \theta^{0j}p^{i}\right) 
\label{11}
\end{eqnarray}
\begin{eqnarray}
\{x^{'0}, x^{'i}\}_{DB} = \frac{1}{m}(\theta^{0i}p_{0}+\theta^{0k}p_{k}
\frac{p^{i}}{p_{0}})
\label{12}
\end{eqnarray}
\begin{eqnarray}
\{x^{'i}, p^{'}_{0}\}_{DB} = \frac{p^{i}}{p_{0}}      \qquad;
 \qquad\{x^{'i}, p^{'}_{j}\}_{DB} = {\delta^{i}}_{j} 
\label{13}
\end{eqnarray}
It is now important to observe that the noncommutativity in
the space-space coordinates (\ref{11}) has a Lie-algebraic structure
in phase-space (with the inclusion of identity) 
and not in space-time\footnote{Note that following \cite{madore}, one can 
therefore associate an appropriate ``diamond star product" 
for this in order to compose any pair of phase-space functions.}. 
This is in contrast to the results derived in \cite{sg}
(for the relativistic free particle)
where space-space noncommutativity was not Lie-algebraic 
in form because of the presence of $p_{0}$ in the denominator
which did not have a vanishing bracket with all other phase-space
variables. However, the algebra between the space-time coordinates is not
Lie-algebraic in form. 

\noindent Alternatively, one may demand that the space-space algebra
between the primed coordinates is of the Lie-algebraic form
(\ref{11}). A simple inspection (after the substitution
of (\ref{10}) in the left hand side of (\ref{11})) gives
the solution (\ref{10a}) for the reparametrisation parameter
$\epsilon$.
\noindent The change of variables relating the primed coordinates with
the unprimed ones hence read:
\begin{eqnarray}
x^{'0} &=& x^{0} - \theta^{0k}p_{k}\frac{p_{0}}{m} \nonumber\\
x^{'i} &=& x^{i} + \theta^{0k}p_{k}\frac{p^{i}}{m} 
\label{140n}
\end{eqnarray}
Note that the above change of variables is different to that derived
in \cite{sg}. This is because $\epsilon$ in \cite{sg} is of the form
$\epsilon=-\theta^{0k}p_{k}$.

\noindent The above solution of $\epsilon$ (\ref{10a})
shows that the desired
gauge fixing condition is given by \footnote{Note that we have dropped
the prime from $x^{\prime}_{0}$ for convenience in (\ref{14y}).}:
\begin{eqnarray}
\phi_{3} = x^{0} + \theta^{0k}p_{k}\frac{p_{0}}{m} - \tau \approx 0,
 \qquad k = 1, 2, ...d.
\label{14y}
\end{eqnarray}
It is easy to check that the constraints (\ref{6}, \ref{14y}) form
a second class pair as
\begin{eqnarray}
\{\phi_{a}, \phi_{b}\} = 2p_{0}\epsilon_{ab};\quad(a,b=1,3). 
\label{10aaa}
\end{eqnarray}
The set of non-vanishing DB(s) consistent with the strong imposition
of the constraints (\ref{6}, \ref{14y}) reproduces the results
(\ref{11}, \ref{12}, \ref{13}).

\noindent We now investigate 
the algebra of the Lorentz generators
(rotations and boosts) for the above choice of the
reparametrisation parameter $\epsilon$ to illuminate the internal consistency
of our analysis. 
%%%%%%%%%%%%%%%%%%%%%%%%%%%%%%%%%%%%%%%%%%
As we have pointed out in \cite{sg}, the
definition of the Lorentz generators remains unchanged in
our approach, because these are gauge invariant objects.
The Lorentz generators (rotations and boosts) are defined as,
\begin{eqnarray}
M_{ij} = x_{i}p_{j} - x_{j}p_{i} 
\label{A1}
\end{eqnarray}
\begin{eqnarray}
M_{0i} = x_{0}p_{i} - x_{i}p_{0} 
\label{A2}
\end{eqnarray}
They satisfy the usual algebra in both the unprimed and the primed
coordinates as $M_{\mu\nu}$ and $p_{\mu}$ are both gauge invariant
\cite{sg}.
\begin{eqnarray}
\{M_{ij}, p_{k}\}_{DB} = \delta_{ik}p_{j} - \delta_{jk}p_{i} 
\label{29.1}
\end{eqnarray}
\begin{eqnarray}
\{M_{ij}, M_{kl}\}_{DB} = \delta_{ik}M_{jl} - \delta_{jk}M_{il} 
+ \delta_{jl}M_{ik} - \delta_{il}M_{jk}
\label{29.2}
\end{eqnarray}
\begin{eqnarray}
\{M_{ij}, M_{0k}\}_{DB} = \delta_{ik}M_{0j} - \delta_{jk}M_{0i} 
\label{29.3}
\end{eqnarray}
\begin{eqnarray}
\{M_{0i}, M_{0j}\}_{DB} = M_{ji} 
\label{29.4}
\end{eqnarray}
\begin{eqnarray}
\{M_{0i}, p_{k}\}_{DB} = -\delta_{ik}p_{0} 
\label{29.5}
\end{eqnarray}

\noindent On the other hand, the algebra between the space coordinates
and those of Lorentz generators (rotations and boosts) 
are different in the two gauges (\ref{9.1}, \ref{14y})
since $x^{k}$ is not gauge invariant under gauge transformation.
We find:
\begin{eqnarray}
\{M_{ij}, x^{k}\}_{DB} = {\delta_{i}}^{k}x_{j} - {\delta_{j}}^{k}x_{i} 
\label{A3}
\end{eqnarray}
\begin{eqnarray}
\{M_{0i}, x^{j}\}_{DB} = x_{i}\frac{p^{j}}{p_{0}} - x_{0}{\delta_{i}}^{j} 
\label{A4}
\end{eqnarray}
\begin{eqnarray}
\{M_{ij}, x^{'k}\}_{DB} 
&=& {\delta_{i}}^{k}x^{'}_{j} - {\delta_{j}}^{k}x^{'}_{i}
 + \frac{p^{k}}{m}\left({\theta^{0}}_{i}p_{j} -
 {\theta^{0}}_{j}p_{i}\right)\nonumber\\ 
\label{A5}
\end{eqnarray}
\begin{eqnarray}
\{M_{0i}, x^{'j}\}_{DB} &=& x^{'}_{i}\frac{p^{j}}{p_{0}}
 - x^{'}_{0}{\delta_{i}}^{j} - {\theta^{0}}_{i}p^{j}\frac{p_{0}}{m}
-\theta^{0k}p_{k}\frac{p_{i}p^{j}}{mp_{0}}  
\label{A6}
\end{eqnarray}
Note at this stage that the gauge choice (\ref{14y}) is not Lorentz invariant.
However, the Dirac bracket procedure forces this constraint equation
to be strongly valid in all Lorentz frames \cite{hanson}. 
This can be made consistent if and only if an infinitesimal Lorentz boost
to a new frame 
\begin{eqnarray}
p^{\mu} \rightarrow p^{'\mu} = p^{\mu} + \omega^{\mu\nu}p_{\nu};\quad
\omega^{\mu\nu}=-\omega^{\nu\mu}
\label{A7a}
\end{eqnarray}
is accompanied by a compensating infinitesimal gauge transformation
\begin{eqnarray}
\tau \rightarrow \tau^{'} = \tau + \Delta\tau
\label{A8}
\end{eqnarray}
The change in $x^{\mu}$, upto first order in $\omega$, is therefore
\begin{eqnarray}
x^{'\mu}(\tau) & = &x^{\mu}(\tau^{'}) + \omega^{\mu\nu}x_{\nu}(\tau)\nonumber\\
 & = & x^{\mu}(\tau) + \Delta\tau\frac{dx^{\mu}}{d\tau} + \omega^{\mu\nu}x_{\nu}
\label{A9}
\end{eqnarray}
In particular, the zero-th component is given by,
\begin{eqnarray}
x^{'0}(\tau) = x^{0}(\tau) + \Delta\tau\frac{dx^{0}}{d\tau} + \omega^{0i}x_{i}
\label{A10}
\end{eqnarray}
Since the gauge condition (\ref{14y}) is $x^{0}(\tau) \approx \tau - 
\theta^{0k}p_{k}\frac{p_{0}}{m}$, $x^{\prime 0}(\tau)$ also must satisfy
$x^{\prime 0}(\tau) = (\tau - \theta^{0k}p^{\prime}_{k}
\frac{p^{\prime}_{0}}{m})$
in the boosted frame, which can now be written, using (\ref{A7a}) as:
\begin{eqnarray}
x^{'0}(\tau) &=& \tau - \frac{\theta^{0k}}{m}(p_{k}+{\omega_{k}}^{0}p_{0})
(p_{0}+{\omega_{0}}^{l}p_{l}).
\label{A11}
\end{eqnarray}
Comparing the left hand side of the above equation with (\ref{A10})
and using the gauge condition (\ref{14y}),
one can now solve for $\Delta \tau$ upto terms linear in
$\omega$ to get,
\begin{eqnarray}
\Delta\tau = \frac{\theta^{0k}}{m}({\omega^{0}}_{k}p_{0}^2-{\omega_{0}}^{l}
p_{l}p_{k})\label{A12u}
\end{eqnarray}
Therefore, for a pure boost, the spatial components of (\ref{A9}) satisfy
\begin{eqnarray}
\delta x^{j}(\tau) &=& x^{'j}(\tau) - x^{j}(\tau) = 
\Delta\tau\frac{dx^{j}}{d\tau} + \omega^{j0}x_{0}\nonumber\\
&=& \omega^{0i}\left(x_{i}\frac{p^{j}}{p_{0}}
 - x_{0}{\delta_{i}}^{j} - {\theta^{0}}_{i}p^{j}\frac{p_{0}}{m}
-\theta^{0k}p_{k}\frac{p_{i}p^{j}}{mp_{0}}\right)+O(\omega^2)
\label{A12a}
\end{eqnarray}
Hence we find that (\ref{A6}) and (\ref{A12a}) are consistent with each other
upto first order in $\omega$.

%%%%%%%%%%%%%%%%%%%%%%%%%%%%%%%%%%%%%%%%%%
\noindent Next we observe that there
is another interesting choice of $\epsilon$ which reads the following: 
\begin{eqnarray}
\epsilon=-d_{k}\theta^{kl}p_{l}\frac{p_{0}}{m}
\label{15aaa}
\end{eqnarray}
where; $d_{k}$ are arbitrary dimensionless constants.

\noindent This yields (using (\ref{9.3}) and (\ref{10}))
the following algebra between the primed coordinates:
\begin{eqnarray}
\{x^{'i}, x^{'j}\}_{DB} = \frac{d_{k}}{m}\left(\theta^{ki}p^{j}
- \theta^{kj}p^{i}\right)
\label{16aaa}
\end{eqnarray}

\begin{eqnarray}
\{x^{'0}, x^{'i}\}_{DB} = \frac{d_{k}}{m}\left(\theta^{ki}p_{0}+
\theta^{kl}p_{l}\frac{p_{i}}{p_{0}}\right)
\label{12aaa}
\end{eqnarray}
\begin{eqnarray}
\{x^{'i}, p^{'}_{0}\}_{DB} = \frac{p_{i}}{p_{0}}      \qquad;
 \qquad\{x^{'i}, p^{'}_{j}\}_{DB} = {\delta^{i}}_{j} 
\label{13aaa}
\end{eqnarray}
Once again we obtain a Lie-algebraic noncommutative structure in the
space-space sector. However, note that (\ref{16aaa}) is different from
(\ref{11}) because the noncommutative parameter $\theta$ in
(\ref{16aaa}) has space indices in contrast to the space-time indices
appearing in (\ref{11}). The space-time algebra is once again
not Lie-algebraic in form.

\noindent The desired gauge fixing condition which leads to the above
Dirac brackets read:
\begin{eqnarray}
\phi_{4} = x^{0} + d_{k}\theta^{kl}p_{l}\frac{p_{0}}{m} - \tau \approx 0,
 \qquad k = 1, 2, ...d.
\label{14}
\end{eqnarray}

\noindent As before the algebra between the Lorentz generators $M_{\mu\nu}$
and $p_{\mu}$ remains the same (\ref{29.1}, \ref{29.2}, \ref{29.3}, 
\ref{29.4}, \ref{29.5}). Also the algebra between
$M_{\mu\nu}$ and $x^{\mu}$ in the unprimed coordinates remain the same
(\ref{A3}, \ref{A4}). However, the algebra between
$M_{\mu\nu}$ and $x^{\prime\mu}$ in the primed coordinates are different
and read:
\begin{eqnarray}
\{M_{ij}, x^{'k}\}_{DB} 
&=& {\delta_{i}}^{k}x^{'}_{j} - {\delta_{j}}^{k}x^{'}_{i}
 + \frac{d_{l}}{m}\left({\theta^{l}}_{i}p_{j} -
 {\theta^{l}}_{j}p_{i}\right)p^{k}\nonumber\\ 
\label{A5aa}
\end{eqnarray}
\begin{eqnarray}
\{M_{0i}, x^{'j}\}_{DB} &=& x^{'}_{i}\frac{p^{j}}{p_{0}}
 - x^{'}_{0}{\delta_{i}}^{j} - d_{l}{\theta^{l}}_{i}p^{j}\frac{p_{0}}{m}
-d_{l}\theta^{lk}p_{k}\frac{p_{i}p^{j}}{mp_{0}}\nonumber\\  
\label{A6aa}
\end{eqnarray}
Rerunning our earlier analysis of enforcing the constraint equation
(\ref{14}) to be strongly valid in all Lorentz frames leads to the
following solution for $\Delta \tau$ in (\ref{A9})
upto terms linear in $\omega$:
\begin{eqnarray}
\Delta\tau = \frac{d_{k}\theta^{kl}}{m}({\omega^{0}}_{l}p_{0}^2-
{\omega_{0}}^{r}p_{l}p_{r}){\omega_{0}}^{r}p_{0}p_{r}-\omega^{0i}x_{i}
\label{A12}
\end{eqnarray}
Therefore, for a pure boost, the spatial components of (\ref{A9}) satisfy
\begin{eqnarray}
\delta x^{j}(\tau) &=& x^{'j}(\tau) - x^{j}(\tau) = 
\Delta\tau\frac{dx^{j}}{d\tau} + \omega^{j0}x_{0}\nonumber\\
&=& \omega^{0i}\left(x_{i}\frac{p^{j}}{p_{0}}
 - x_{0}{\delta_{i}}^{j} - d_{k}{\theta^{k}}_{i}p^{j}\frac{p_{0}}{m}
-d_{k}\theta^{kl}p_{l}\frac{p_{i}p^{j}}{mp_{0}}\right)+O(\omega^2)
\label{A12aa}
\end{eqnarray}
Hence we find that (\ref{A6aa}) and (\ref{A12aa}) are consistent with
each other upto first order in $\omega$.
%%%%%%%%%%%%%%%%%%%%%%%%%%%%%%%%%%%%%%%%%%%%%%%%%%%%%%%%%%%%%%%%%%%

%%%%%%%%%%%%%%%%%%%%% Observations %%%%%%%%%%%%%%%%%%%%%%%%%%%%%%%%
\noindent We make certain observations now. Although, the relations 
(\ref{11}), (\ref{16aaa})
have a close resemblance to Snyder's algebra \cite{snyder}, there is a 
subtle difference. Note that the right hand side of these relations
do not have the structure of an angular momentum
operator in their differential representation
(obtained by repacing $p_{j}$ by ($-i\partial_{j}$) in contrast 
to the Snyder's algebra. Further, the relations are not
reminiscent of $\kappa$-Minkowski algebra 
(that has been studied extensively in the literature recently
\cite{amelino}, \cite{lukierski}, \cite{kowalski}, \cite{subir})
but has a similar structure to the commutation relations
describing the Lie-algebraic deformation of the
Minkowski space \cite{woro}, 
the only difference being that
momentum operators appear at the right hand side of the relations 
instead of the position operators.
\noindent Interestingly, the values of the
reparametrisation parameter $\epsilon$ (\ref{10a}, \ref{15aaa})
that leads to the noncommuting Lie-algebraic structures 
(\ref{11}, \ref{16aaa}) are
the only choices possible as can be easily seen from purely
dimensional considerations. 
%%%%%%%%%%%%%%%%%%%%%%%%%%%%%%%%%%%%%%%%%%%%%%%%%%%%%%%%%%%%%%%%%%%

%%Other choices%%%%%%%%%%%%%%%%%%%%%%%%%%%%%%%%%%%%%%%%%%%%%%%%%%%%
\noindent Finally, there exists choices of $\epsilon$
for which the space-space noncommutativity
can be made to vanish. The choices are:
\begin{eqnarray}
\epsilon=e_{k}\theta^{0k}\frac{p_{0}^2}{m}
\label{14a}
\end{eqnarray}
and
\begin{eqnarray}
\epsilon=-f_{kl}\theta^{kl}p_{0}
\label{14b}
\end{eqnarray}
where, $e_{k}$ and $f_{kl}$ are arbitrary dimensionless constants.

\noindent The space-time algebras however
do not vanish for the above values of
$\epsilon$ and are as follows:
\begin{eqnarray}
\{x^{\prime 0}, x^{\prime i}\}=\frac{2e_{k}}{m}\theta^{0k}p^{i}
\label{14c}
\end{eqnarray}
and
\begin{eqnarray}
\{x^{\prime 0}, x^{\prime i}\}=f_{kl}\theta^{kl}\frac{p^{i}}{p_{0}}.
\label{14d}
\end{eqnarray}

%%%%%%%%%%%%%%%%%%%%%%%%%%%%%%%%%%%%%%%%%%%%%%%%%%%%%%%%%%
%%%%%%%%%%%%%%%%%%%%%%%%%%%%%%%%%%%%%%%%%%%%%%%%%%%%%%%%%%
\noindent Another interesting question that can be asked is the 
following. Can we get canonical NC space-space structures
from reparametrisation symmetry. 
To see this we ask whether there exists a solution of
$\epsilon$ for which the following equation holds
between the primed coordinates:
\begin{eqnarray}
\{x^{\prime i}, x^{\prime j}\}=\theta^{ij}.
\label{1.1}
\end{eqnarray}
Substituting (\ref{10}) in the left hand side of the above equation in
terms of the unprimed coordinates, it is easy to note that there
does not exist an $\epsilon$ for which
the above equation is satisfied. Thus, one cannot obtain canonical
NC space-space structures from reparametrisation symmetry.

%----------------------------------------------------------------
\section{Conclusions}
In this paper, we have extended our earlier work on
noncommutativity and reparametrisation symmetry \cite{sg}
to obtain Lie-algebraic NC structures in case of a free
relativistic particle. This is in contrast to the results
obtained in \cite{sg} since the NC structures
although Lie-algebraic in form in case of the non-relativistic
free particle were not so in its relativistic counterpart.
The change of variables derived in this paper are
different than those appearing in \cite{sg}. This is related to the
fact that the choice of the reparametrisation parameter $\epsilon$
is different in the two cases.

\noindent We also find solutions (for $\epsilon$) for which
the algebra between space-space coordinates in the primed
sector vanishes while the space-time algebra still survives.
However, we do not get canonical (\ref{1}) or quantum space structures
(\ref{3}) from reparametrisation symmetry.

\noindent As a matter of internal consistency of our analysis,
we study the angular momentum algebra in details. As in
\cite{sg}, the angular momentum remains gauge invariant since the change
of variables is just a gauge transformation. Hence, we feel that
our approach is more elegant than those \cite{stern} where such
change of variables are found by inspection and leads to ambiguities
in the definition of physical variables like the angular momentum.

%----------------------------------------------------------------
%**********************************
\section*{Acknowledgement}
It is a pleasure to thank Dr.Biswajit Chakraborty for a careful
reading of the manuscript and for his questions and comments
on it. The author would also like to thank the referee for very
useful comments.
%**********************************

%----------------------------------------

\end{document}